\documentclass{ws-procs975x65}

\begin{document}

\title{\uppercase{Primordial Gravitational Waves and Cosmic Microwave Background Radiation}}

\author{\uppercase{Deepak~Baskaran$^{1,2*}$, Leonid~P.~Grishchuk$^{1,3}$, Wen~Zhao$^{1,2,4}$}}

\address{$^1$School of Physics \& Astronomy, Cardiff University, Cardiff, CF24 3AA, UK,\\  
$^2$Wales Institute of Mathematical \& Computational Sciences, Swansea, SA2 8PP, UK,\\
$^3$Sternberg Astronomical Insititute, Moscow State University, Moscow, 119899, Russia,\\
$^4$Department of Physics, Zhejiang University of Technology, Hangzhou, 310014, People's 
Republic of China.\\
$^*$E-mail: Deepak.Baskran@astro.cf.ac.uk}

\begin{abstract}
This is a summary of presentations delivered at the OC1 parallel session 
``Primordial Gravitational Waves and the CMB" of the $12^{\rm th}$ Marcel Grossmann 
meeting in Paris, July 2009. The reports and discussions demonstrated significant progress 
that was achieved in theory and observations. It appears that the existing data provide 
some indications of the presence of gravitational wave contribution to the CMB anisotropies, 
while ongoing and planned observational efforts are likely to convert these indications 
into more confident statements about the actual detection.  
\end{abstract}


\bodymatter

%%%%%%%%%%%%%%%%%%%%%%%%%%%%%%%%%%%%%%%%%%%%%%%%%%%%%%%%%%%%%%des%%%%%%%%%%%%

\section{Introduction}\label{sec1}

Here, we report on presentations and discussions that took place
at the OC1 parallel session entitled ``Primordial Gravitational Waves 
and the CMB". The programme was designed to include both theoretical 
and observational results. The electronic versions of some of the 
talks delivered at this session are available at the MG12 website \cite{mg12website}. 
We shall start from the overall framework of this session and motivations 
for studying the primordial (relic) gravitational waves. 

Primordial gravitational waves (PGWs) are necessarily generated by a strong 
variable gravitational field of the very early Universe \cite{gr74}.
The existence of relic grtavitational waves relies only on the validity of 
basic laws of general relativity and quantum mechanics. Specifically, the 
generating mechanism is the superadiabatic (parametric) amplification of the waves' 
zero-point quantum oscillations. In contrast to other known massless particles, 
the coupling of gravitational waves to the external (``pump") gravitational 
field is such that they could be classically amplified or quantum-mechanically 
generated by the gravitational field of a homogeneous isotropic FLRW 
(Friedmann-Lemaitre-Robertson-Walker) universe. Under certain extra conditions
the same applies to the primordial density perturbations. The PGWs are the 
cleanest probe of the physical conditions in the early Universe right down 
to the limits of applicability of currently available theories, i.e. the Planck 
density $\rho_{\rm Pl} = c^5/G^2 \hbar \approx 10^{94}{\rm g}/{\rm cm}^{3}$ 
and the Planck size $l_{\rm Pl} = (G \hbar/c^3)^{1/2} \approx 10^{-33}{\rm cm}$. 

The amount and spectral content of the PGWs field depend on the evolution of the 
cosmological scale factor $a(\eta)$ representing the gravitational pump field. 
The theory was applied to a variety of $a(\eta)$, including 
those that are now called inflationary models \cite{gr74,a12a,starobinski}. If the 
exact $a(\eta)$ were known in advance from some fundamental ``theory-of-everything", 
we would have derived the properties of the today's signal with no ambiguity. In 
the absence of such a theory, we have to use the available partial information 
in order to reduce the number of options. The prize is very high - the actual detection 
of a particular background of PGWs will provide us with a unique clue to the 
birth of the Universe and its very early dynamical behaviour. 

To be more specific, let us put PGWs in the context of a complete cosmological 
theory hypothesizing that the observed Universe has come to the existence with  
near-Planckian energy density and size (see papers \cite{birthpapers,Grishchuk2009} and 
references therein). It seems reasonable to conjecture that the embryo Universe was 
created by a quantum-gravity or by a ``theory-of-everything" process in a 
near-Planckian state and then started to expand. (If you think that the development 
of a big universe from a tiny embryo is arrant nonsense, you should recollect that 
you have also developed from a single cell of microscopic size. Analogy proposed by 
the biophysicist E. Grishchuk.) The total energy, including gravity, of the 
emerging classical configuration was likely to be zero then and remains zero now. 

\begin{figure}
\includegraphics[width=12cm,height=10cm]{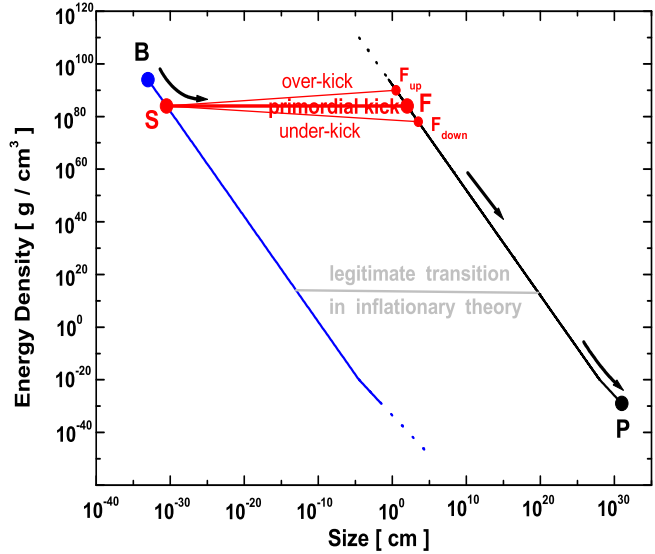}
\caption{A primordial kick is required in order to reach the 
present state of the Universe P from the birth event B. Red lines describe
possible transitions that would be accompanied by the generated cosmological 
perturbations of observationally necessary level and spectral shape  
\cite{Grishchuk2009}. The ``legitimate transition in inflationary theory" is an 
evolution allowed by the incorrect (inflationary) formula for density perturbations. 
See explanations in Sec.{\ref{section3}} below.}
\label{birth}  
\end{figure}

In order for the natural hypothesis of spontaneous birth of the observed Universe 
to bring us anywhere near our present state 
characterized by $\rho_p =3 H_0^2/8 \pi G \approx 10^{-29} {\rm g}/{\rm cm}^3$ 
and $l_p = 500 l_H$ (which is the minimum size of the present-day patch of homogeneity 
and isotropy, as follows from observations \cite{GrZeld}) the newly-born Universe 
needs a significant `primordial kick'. During the kick, the size of the 
Universe (or, better to say, the size of our patch of homogeneity and isotropy) 
should increase by about 33 orders of magnitude without losing too much of the
energy density of whatever substance that was there, or maybe even slightly increasing
this energy density at the expence of the energy density of the gravitational field.
This process is graphically depicted in Fig.~\ref{birth} (adopted from the paper 
\cite{Grishchuk2009}). The present state of the accessible Universe is marked by the 
point P, the birth of the Universe is marked by the point B. If the 
configuration starts at the point B and then expands according to the usual laws 
of radiation-dominated and matter-dominated evolution (blue curve), it completely 
misses the desired point P. By the time the Universe has reached the size $l_p$,
the energy density of its matter content would have dropped to the level many orders of 
magnitude lower than the required $\rho_p$. The only way to reach P from B is to assume 
that the newly-born Universe has experienced a primordial kick allowing the point of 
evolution to jump over from the blue curve to the black curve. 

If we were interested only in the zero-order approximation of homogeneity and isotropy, 
there would be many evolutionary paths equally good for connecting the points B 
and P. However, in the next-order approximations, which take into account the inevitable 
quantum-mechanical generation of cosmological perturbations, the positioning and form of 
the transition curve in Fig.~\ref{birth} become crucial. The numerical value of 
the Hubble parameter $H$ (related to the energy density of matter driving the kick, as shown
on the vertical axis of the figure) determines the numerical level of amplitudes of the 
generated cosmological perturbations, while the shape of the transition curve determines 
the shape of the primordial power spectrum.

The simplest assumption about the initial kick is that its entire duration was 
characterized by a single power-law scale factor \cite{gr74}
\begin{eqnarray}
\label{scfactor}
a(\eta) = l_o|\eta|^{1+\beta},
\end{eqnarray}
where $l_o$ and $\beta$ are constants, and $\beta < -1$. In this power-law case, the 
gravitational pump field is such that the generated primordial metric power spectra 
(primordial means considered for wavelengths longer than the Hubble radius at the 
given moment of time), for both gravitational waves and density perturbations, have the 
universal power-law dependence on the wavenumber $n$:
\begin{equation}
\label{primsp}
h^{2}(n) \propto n^{2(\beta+2)}.
\end{equation}
It is common to write these metric power spectra separately for gravitational waves (gw) 
and density perturbations (dp): 
\begin{equation}
\label{primsp1}
h^2(n)~({\rm gw}) = B_t^2 n^{n_t}, ~~~~~h^2(n)~({\rm dp})=B_s^2 n^{n_s -1}.
\end{equation}

According to the theory of quantum-mechanical generation of cosmological 
perturbations (for a review, see \cite{a12a}), the spectral indices are approximately 
equal, $n_s-1 = n_t = 2(\beta+2)$, and the amplitudes $B_t$ and $B_s$ are of the 
order of magnitude of the ratio $H_i/H_{\rm Pl}$, where $H_i\sim c/l_o$ is the 
characteristic value of the Hubble parameter $H$ during the kick.  
The straight lines in Fig.~\ref{birth} symbolize the power-law kicks 
(\ref{scfactor}). They generate primordial spectra with constant spectral indices 
throughout all wavelenghts. In particular, any horizontal line describes an 
interval of de Sitter evolution, $\beta = -2$, ${\dot H} =0$, $H = const$. (Initial 
kick driven by a scalar field is appropriately called inflation: 
dramatic increase in size with no real change in purchasing power, i.e. in matter 
energy density.) The gravitational pump field of a de Sitter kick transition 
generates perturbations with flat (scale-independent) spectra $n_s-1 = n_t = 0$. 
The red horizontal line shown in Fig.~\ref{birth} corresponds to 
$H_i/H_{\rm Pl} \approx 10^{-5}$  and the generated primordial amplitudes 
$B_t \approx B_s \approx 10^{-5}$. In numerical calculations,
the primordial metric power spectra are usually parameterized by 
\begin{eqnarray}
\label{PsPt}
P_{t}(k)=A_t(k/k_0)^{n_t},~~P_{s}(k)=A_s(k/k_0)^{n_s-1},
\end{eqnarray}
where $k_0=0.002$Mpc$^{-1}$. 

Although the assumption of a single piece of power-law evolution (\ref{scfactor}) 
is simple and easy to analyze, the reality could be more complicated and probably 
was more complicated (see the discussion of WMAP data in Sec.\ref{section3}). A less
simplistic kick can be approximated by a sequence of power-law evolutions, and 
then the primordial power spectra will consist of a sequence of power-law intervals.  

The amplitudes of generated cosmological perturbations are large at 
long wavelengths. According to the widely accepted assumption the observed
anisotropies in the cosmic microwave background radiation (CMB) are caused
by perturbations of quantum-mechanical origin. This assumption is
partially supported by the observed ``peak and dip" structure of the CMB angular
power spectra. Presumably, this structure reflects the phenomenon of 
quantum-mechanical phase squeezing and standing-wave pattern of the generated metric 
fields \cite{a12a}. The search for the relic gravitational waves is a goal of 
a number of current and future space-borne, sub-orbital and ground-based CMB 
experiments \cite{Planck,WMAP5,BICEP,quad,Clover,QUITE,EBEX,SPIDER,cmbpol}. 

The CMB anisotropies are usually characterized by the four angular 
power spectra $C_{\ell}^{TT}$, $C_{\ell}^{EE}$, $C_{\ell}^{BB}$ and 
$C_{\ell}^{TE}$ as functions of the multipole number $\ell$. The contribution 
of gravitational waves to these power spectra has been studied, both 
analytically and numerically, in a number of papers 
\cite{Polnarev1,a8,a11,a12,a13}. The derivation of today's CMB 
power spectra brings us to approximate formulas of the following structure \cite{a12}:
\begin{eqnarray}
\label{exact-clxx'} 
\begin{array}{l} 
C_{\ell}^{TT}= \int \frac{dn}{n} [h(n, \eta_{rec})]^2 \left[F^T_{\ell}(n)\right]^2, \\
C_{\ell}^{TE}=\int \frac{dn}{n} h(n, \eta_{rec}) h^{\prime}(n, \eta_{rec}) 
\left[F^T_{\ell}(n) F^E_{\ell}(n)\right], \\
C_{\ell}^{YY}=\int \frac{dn}{n} [h^{\prime} (n, \eta_{rec})]^2 
\left[F^Y_{\ell}(n)\right]^2, ~~~~{\rm where}~Y=E,B.
\end{array}
\end{eqnarray}
In the above expressions, $[h(n, \eta_{rec})]^2$ and $[h^{\prime} (n, \eta_{rec})]^2$
are power spectra of the gravitational wave field and its first 
time-derivative. The spectra are taken at the recombination (decoupling) time 
$\eta_{rec}$. The functions $F^X_{\ell}(n)$ ($X=T, E, B$) take care of the 
radiative transfer of CMB photons in the presence of metric perturbations. To a good 
approximation, the power residing in the metric fluctuations at wavenumber $n$ 
translates into the CMB $TT$ power at the angular scales characterized by the 
multipoles $\ell \approx n$. Similar results hold for the CMB power spectra 
induced by density perturbations. Actual numerical calculations use equations more accurate
than Eq.~(\ref{exact-clxx'}).

There are several differences between the CMB signals arising from density 
perturbations and gravitational waves. For example, gravitational waves 
produce B-component of polarization, while density perturbations do not \cite{a8}; 
gravitational waves produce negative TE-correlation at lower multipoles, while
density perturbations produce positive correlation 
\cite{a12a,a12,polnarev,zbg,zbg2}, and so on. However, it is important to realize that 
it is not simply the difference between zero and non-zero or between positive and 
negative that matters. (In any case, various noises and systematics will make 
this division much less clear cut.) What really matters is that the gw and dp sources 
produce different CMB outcomes, and they are in principle distinguishable, even if 
they both are non-zero and of the same sign. For example, if the parameters of density 
perturbations were precisely known from other observations, any observed deviation 
from the expected $TT$, $TE$ and $EE$ correlation functions could be attributed 
(in conditions of negligible noises) to gravitational waves. From this perspective, 
the identification of the PGWs signal goes well beyond the often stated goal of 
detecting the B-mode of polarization. In fact, as was argued in the talk by 
D.~Baskaran, delivered on behalf of the group including also L.~P.~Grishchuk 
and W.~Zhao, the $TT$, $TE$, and $EE$ observational channels could be much more 
informative than the $BB$ channel alone. Specifically for the Planck mission, the 
inclusion of other correlation functions, in addition to $BB$, will significantly 
increase the expected signal-to-noise ratio in the PGWs detection.

It is convenient to compare the gravitational wave signal in CMB with that 
induced by density perturbation. A useful measure, directly related to 
observations, is the quadrupole ratio $R$ defined by
\begin{eqnarray}
\label{defineR}
R \equiv \frac{C_{\ell=2}^{TT}({\rm gw})}{C_{\ell=2}^{TT}({\rm dp})},
\end{eqnarray}
i.e. the ratio of contributions of gw and dp to the CMB temperature quadrupole. 
Another measure is the so-called tensor-to-scalar ratio $r$. This quantity is 
built from primordial power spectra (\ref{PsPt}):
\begin{eqnarray}
\label{definer}
 r \equiv \frac{A_t}{A_s}.
\end{eqnarray}
Usually, one finds this parameter linked to incorrect (inflationary) statements. 
Concretely, inflationary theory substitutes (for ``consistency") its prediction of 
arbitrarily large amplitudes of density perturbations in the limit of models where 
spectral index $n_s$ approaches $n_s=1$ and $n_t$ approaches $n_t=0$ (horizontal 
transition lines in Fig.~\ref{birth}) by the claim that it is the amount of relic 
gravitational waves, expressed in terms of $r$, that should be arbitrarily small. 
(For more details, see Sec.~{\ref{section3}}.) However, if $r$ is defined by 
Eq.(\ref{definer}) without implying inflationary claims, one can use 
this parameter. Moreover, one can derive a relation between $R$ and $r$ which depends 
on the background cosmological model and spectral indices. For a rough comparison 
of results one can use $r\approx2R$.

\section{Overview of oral presentations}

\begin{figure}
\begin{center}
\includegraphics[width=12cm]{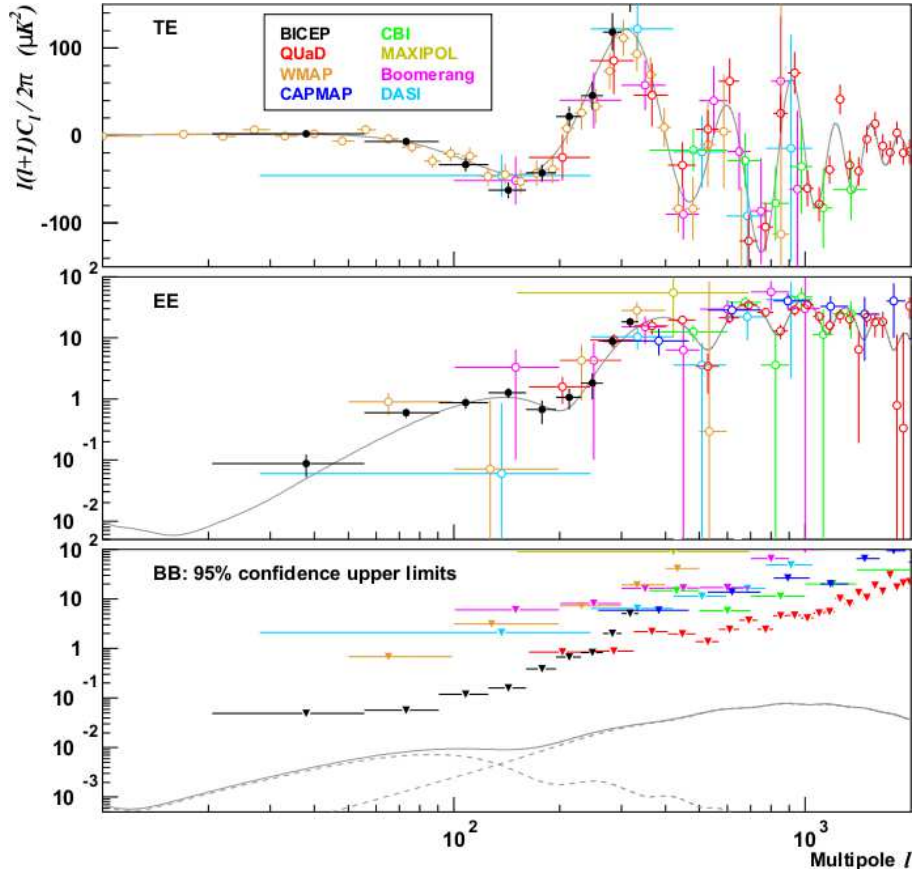}
\end{center}
\caption{BICEPs TE, EE, and BB power spectra together with data from other CMB 
polarization experiments. Theoretical spectra from a CDM model with $r=0.1$ are shown 
for comparison. The BB curve is the sum of the gravitational wave and lensing components. 
At degree angular scales BICEPs constraints on $BB$ are the most powerful to 
date \cite{keating2009}. }
\label{Keatingfig1}
\end{figure}

The OC1 session opened with the talk of Brian Keating, appropriately entitled 
``The Birth Pangs of the Big Bang in the Light of BICEP". The speaker reported on the 
initial results from the Background Imaging of Cosmic Extragalactic Polarization (BICEP) 
experiment. The conclusions were based on data from two years of observation. For
the first time, some meaningful limits on $r$ were set exclusively from the fact of 
(non)observation of the B-mode of polarization (see figure \ref{Keatingfig1} adopted 
from Chiang et al. paper\cite{keating2009}). The author paid attention to various 
systematic effects and showed that they were smaller than the statistical errors. 
B.~Keating explained how the BICEP's design and observational strategies will serve 
as a guide for future experiments aimed at detecting primordial gravitational waves. 
In conclusion, it was stressed that a further $90\%$ increase in the amount of 
analyzed BICEP data is expected in the near future. 

The next talk, delivered by Deepak Baskaran (co-authors: 
L.~P.~Grishchuk and W.~Zhao), was entitled ``Stable indications of relic gravitational 
waves in WMAP data and forecasts for the Planck mission". D. Baskaran reported on 
the results of the likelihood analysis, performed by this group of authors, of the WMAP 
5-year $TT$ and $TE$ data at lower multipoles. Obviously, in the center of their effort
was the search for the presence of a gravitational wave signal \cite{zbg,zbg2}. For the 
parameter $R$, the authors found the maximum likelihood value $R=0.23$, indicating a 
significant amount of gravitational waves. Unfortunately, this determination is 
surrounded by large uncertainties due to remaining noises. This means that the 
hypothesis of no gravitational waves, $R=0$, cannot be excluded yet with any significant 
confidence. The speaker compared these findings with the result of WMAP team, which found
no evidence for gravitational waves. The reasons by which the gw signal can be overlooked
in a data analysis were identified and discussed. Finally, D.~Baskaran presented the 
forecasts for the Planck mission. It was shown that the stable indications of relic 
gravitational waves in the WMAP data are likely to become a certainty in the Planck 
experiment. Specifically, if PGWs are characterized by the maximum likelihood parameters 
found \cite{zbg,zbg2} from WMAP5 data, they will be detected by {\it Planck} at the 
signal-to-noise level $S/N=3.65$, even under unfavorable conditions in terms of 
instrumental noises and foregrounds. (For more details along these lines, see  
sections below.)

The theoretical aspects of generation of CMB polarization and temperature anisotropies by 
relic gravitational waves were reviewed by Alexander Polnarev in the contribution ``CMB 
polarization generated by primordial gravitational waves. Analytical solutions".  
The author described the analytical methods of solving the radiative 
transfer equations in the presence of gravitational waves. This problem is usually tackled 
by numerical codes, but this approach foreshadows the ability of the researcher to 
understand the physical origin of the results. The analytical methods are a useful complement 
to numerical techniques. They allow one not only to explain in terms of the 
underlying physics the existing features of the final product, but also anticipate the 
appearance of new features when the physical conditions change. A. Polnarev showed how the 
problem of CMB anisotropies induced by gravitational waves can be reduced to a single integral 
equation. This equation can be further analyzed in terms of some integral and differential 
operators. Building on this technique, the author formulated analytical solutions as 
expansion series over gravitational wave frequency. He discussed the resonance generation 
of polarization and possible observational consequences of this effect.

An overview of the current experimental situation was delivered by Carrie MacTavish in
the talk ``CMB from space and from a balloon". C. MacTavish focused on the interplay 
between experimental results from two CMB missions: the Planck satellite and the Spider 
balloon experiment. Spider is scheduled for flight over Australia in spring 2010 making
measurements with 1600 detectors. The speaker emphasised the important combined impact 
of these two experiments on determination of cosmological parameters in general.

The ever increasing precision of CMB experiments warrants analysis of subtle observational 
effects inspired by the ideas from particle physics. The talk by Stephon Alexander 
``Can we probe leptogenesis from inflationary primordial birefringent gravitational 
waves" discussed a special mechanism of production of the lepton asymmetry with the 
help of cosmological birefringent gravitational waves. This mechanism was proposed in 
the recent paper \cite{Alexander}. The mechanism assumes that gravitational waves are generated 
in the presence of a CP violating component of the inflaton field that couples to 
a gravitational Chern-Simons term (Chern-Simons gravity). The lepton number
arises via the gravitational anomaly in the lepton number current. As pointed out 
by the speaker, the participating gravitational waves should lead to a unique parity 
violating cross correlation in the CMB. S.~Alexander discussed the viability of 
detecting such a signal, and concluded by analyzing the corresponding observational 
constraints on the proposed mechanism of leptogenesis.

Apart from being an arena for detecting relic gravitational waves, CMB is of course 
a powerful tool for cosmology in general. The final talk of the session, delivered 
by Grant Mathews (co-authors D.~Yamazaki, K.~Ichiki and T.~Kajino), discussed the 
``Evidence for a primordial magnetic field from the CMB temperature and polarization 
power spectra". This is an interesting subject as the magnetic fields are abundant 
in many astrophysical and cosmological phenomena. In particular, 
primordial magnetic fields could manifest themselves in the temperature and 
polarization anisotropies of the CMB. The speaker reported on  
a new theoretical framework for calculating CMB anisotropies 
along with matter power spectrum in the presence of magnetic fields with
power-law spectra. The preliminary evidence from the data on matter and CMB power 
spectra on small angular scales suggest an upper and a lower limit on the
strength of the magnetic field and its spectral index. It was pointed out that this 
determination might be the first direct evidence of the presence of primordial 
magnetic field in the era of recombination. Finally, the author showed that the 
existence of such magnetic field can lead to an independent constraint on the 
neutrino mass.

\section{Analysis of WMAP data and outlook for {\it Planck}}

Along the lines of the presentation by Baskaran et al, we review the results of the 
recent analysis of WMAP 5-year data. The results suggest the evidence, 
although still preliminary, that relic gravitational waves are present, and in the 
amount characterized by $R\approx0.23$, This conclusion follows from the likelihood 
analysis of WMAP5 $TT$ and $TE$ data at lower multipoles $\ell\leq100$. It is
only within this range of multipoles that the power in relic gravitational waves 
is comparable with that in density perturbations, and gravitational waves compete 
with density perturbations in generating CMB temperature and polarization 
anisotropies. At larger multipoles, the dominant signal in CMB comes primarily from 
density perturbations.

\subsection{Likelihood analysis of WMAP data
\label{section2.1}}

The analysis in papers \cite{zbg,zbg2} was based on proper specification of the 
likelihood function. Since $TT$ and $TE$ data are the most informative in the WMAP5, 
the likelihood function was marginalized over the remaining data variables $EE$ 
and $BB$. In what follows $D_{\ell}^{TT}$ and $D_{\ell}^{TE}$
denote the estimators (and actual data) of the $TT$ and $TE$ power spectra. 
The joint pdf for $D_{\ell}^{TT}$ and $D_{\ell}^{TE}$ has the form
\begin{eqnarray}
f(D_{\ell}^{TT},D_{\ell}^{TE})= n^2{x}^{\frac{n-3}{2}}
\left\{2^{1+n}\pi\Gamma^2(\frac{n}{2})(1-\rho_{\ell}^2)(\sigma_{\ell}^T)^{2n}
(\sigma_{\ell}^E)^2\right\}^{-\frac{1}{2}}
\nonumber\\
 \times\exp\left\{\frac{1}{1-\rho^2_{\ell}}\left(\frac{{\rho_{\ell}}
{z}}{{\sigma_\ell^T}{\sigma_\ell^E}}-\frac{{z}^2}{2x{(\sigma_\ell^E)^2}
}-\frac{{x}}{2{(\sigma_\ell^T)^2}}\right)\right\}.
 \label{pdf_CT}
\end{eqnarray}
This pdf contains the variables (actual data) $D_{\ell}^{XX'}$ 
($XX'=TT,TE$) in quantities ${x}\equiv n(D_\ell^{TT}+N_{\ell}^{TT})$ 
and ${z}\equiv nD_\ell^{TE}$, where $N_{\ell}^{TT}$ is the temperature total noise 
power spectrum. The quantity $n= (2\ell+1)f_{\rm sky}$ is the number of effective 
degrees of freedom at multipole $\ell$, where $f_{\rm sky}$ is the sky-cut factor. 
$f_{\rm sky}=0.85$ for WMAP and $f_{\rm sky}=0.65$ for {\it Planck}. $\Gamma$ 
is the $Gamma$-function. The quantities $\sigma_\ell^T$, $\sigma_\ell^E$ and $\rho_\ell$, 
include theoretical power spectra $C_{\ell}^{XX'}$ and contain the 
perturbation parameters $R$,  $A_s$, $n_s$ and $n_t$, whose numerical values the 
likelihood analysis seeks to find.

\begin{figure}
\begin{center}
\includegraphics[width=6cm,height=7cm]{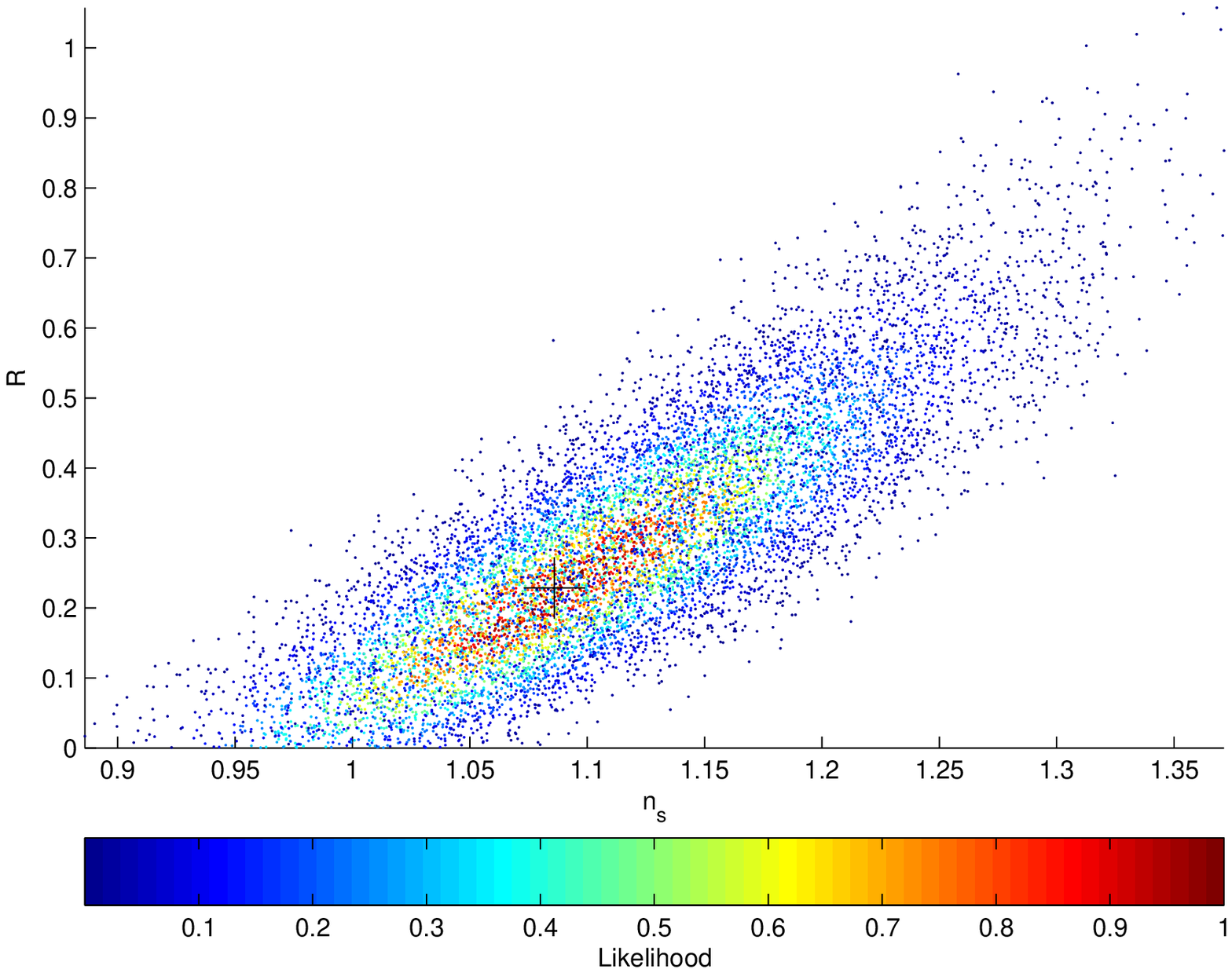}\includegraphics[width=6cm,height=7cm]{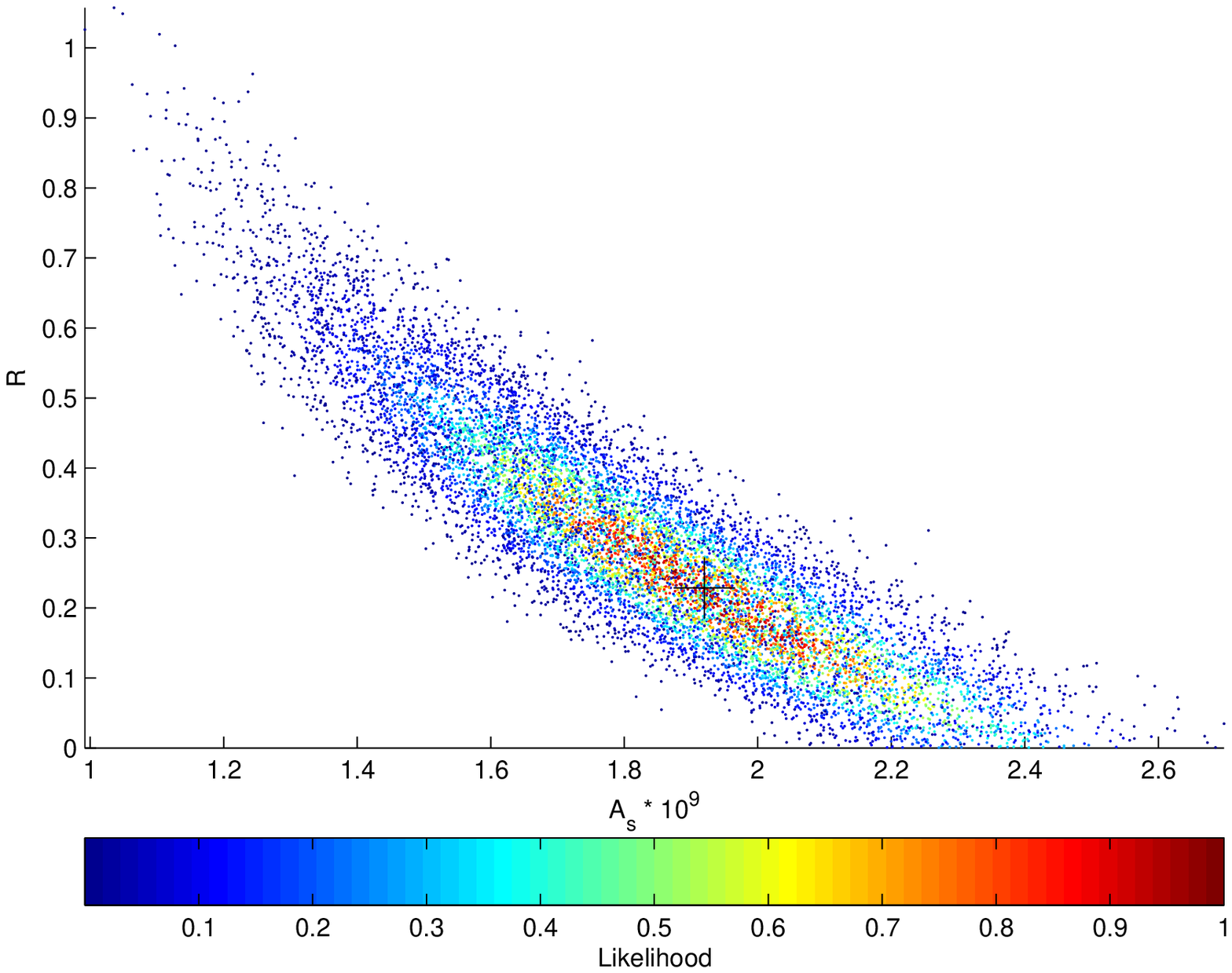}
\end{center}
\caption{The projection of 10,000 samples of the 3-dimensional likelihood 
function onto the  $R-n_s$ (left panel) and $R-A_s$ (right panel) planes. 
The color of an individual point  in Fig.~\ref{figurea1.1} 
signifies the value of the 3-dimensional likelihood of 
the corresponding sample. The black $+$ indicates the parameters listed in (\ref{best-fit}). 
Figure adopted from Zhao et al \cite{zbg2}.}
\label{figurea1.1}
\end{figure}

The three free perturbation parameters $R$, $A_s$, $n_s$ ($n_t=n_s-1$) were 
determined by maximizing the likelihood function
\begin{eqnarray}
\nonumber\label{ctlikelihood1}
 \mathcal{L}\propto \prod_{\ell}f(D_{\ell}^{TT}, D_{\ell}^{TE})
\end{eqnarray}
for $\ell=2...100$. The background cosmological model was fixed at the 
best fit $\Lambda$CDM cosmology \cite{wmap5}. 
The maximum likelihood (ML) values of the perturbation parameters
(i.e.~the best fit values), were found to be
\begin{eqnarray}
R=0.229,~~~n_s=1.086,~~~A_s=1.920\times10^{-9}
\label{best-fit}
\end{eqnarray} 
and $n_t=0.086$. The region of maximum likelihood was probed by 10,000 sample points 
using MCMC method. The projections of all 10,000 points on 2-dimensional 
planes $R-n_s$ and $R-A_s$ are shown in Fig.~\ref{figurea1.1}.  

The samples with relatively large values of the likelihood (red, yellow and green 
colors) are concentrated along the curve, which projects into approximately straight 
lines (at least, up to $R \approx 0.5$): 
\begin{eqnarray}
\label{1Dmodel}
n_s=0.98+0.46R, ~~~~~A_s=(2.27-1.53R)\times10^{-9}.
\end{eqnarray}
These combinations of parameters $R, n_s, A_s$ produce roughly equal 
responses in the CMB power spectra. The best fit model (\ref{best-fit}) is a 
particular point on these lines, $R=0.229$. The family of models (\ref{1Dmodel}) is 
used in Sec.\ref{section4} for setting the expectations for the Planck experiment.  

\begin{figure}
\begin{center}  
\includegraphics[width=6cm,height=6cm]{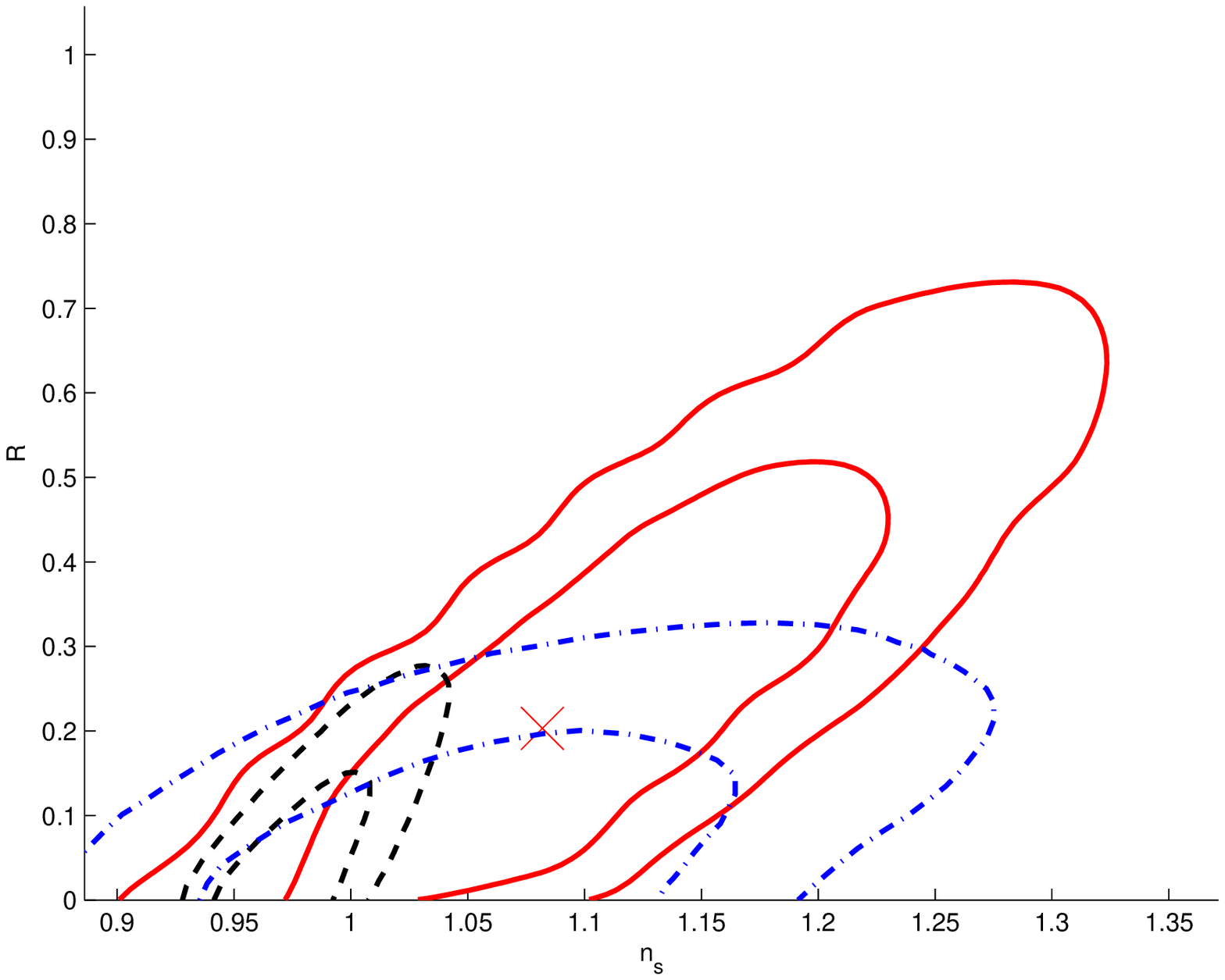}\includegraphics[width=6cm,height=6cm]{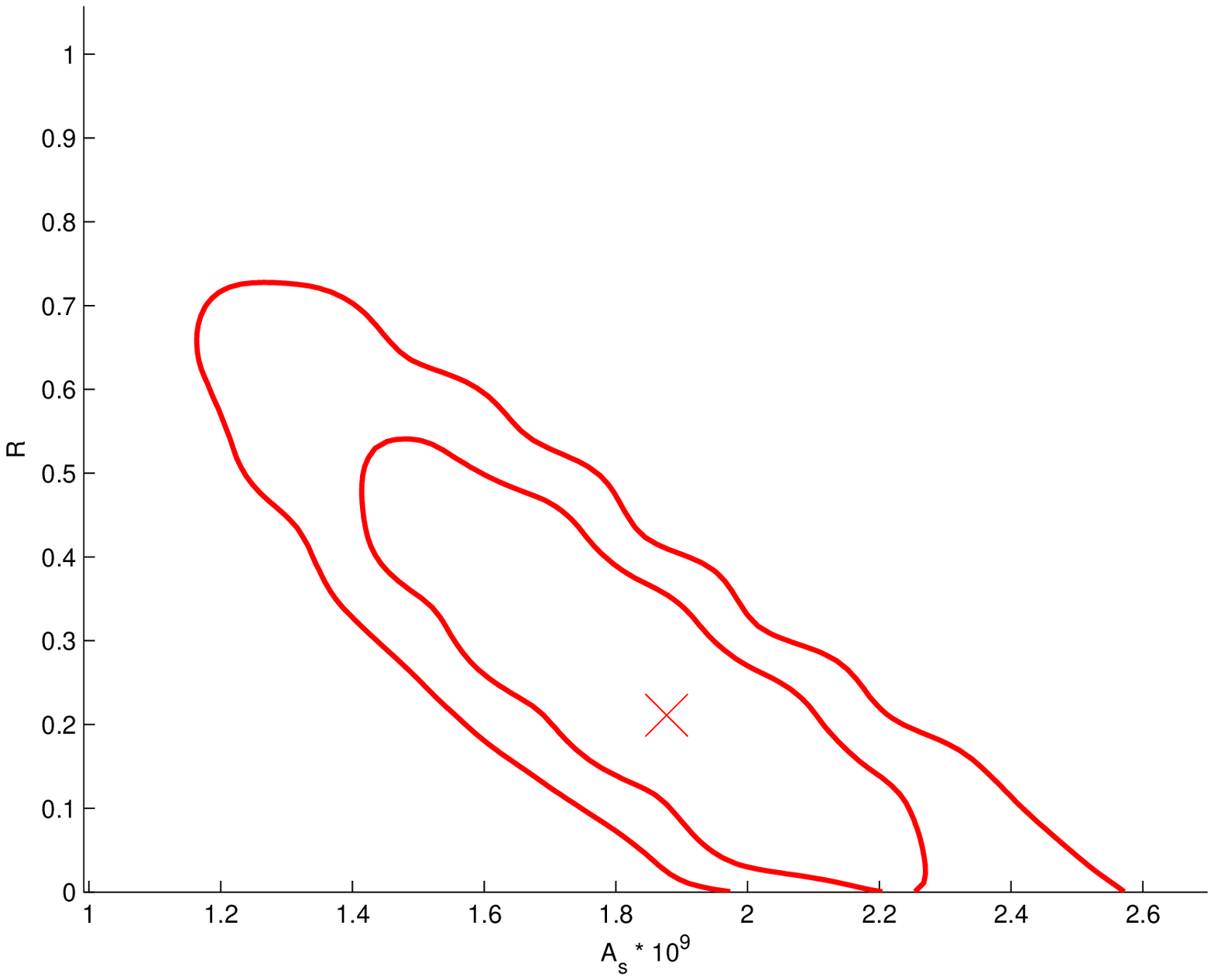}
\end{center}
\caption{The ML points (red $\times$) and the $68.3\%$ and $95.4\%$ confidence 
contours (red solid lines) for 2-dimensional likelihoods: $R-n_s$ (left panel) 
and $R-A_s$ (right panel). In the left panel, the WMAP confidence contours 
are also shown for comparison. Figure adopted from Zhao et al \cite{zbg2}.}
\label{figurea1}
\end{figure}

\begin{figure}
\begin{center}
\includegraphics[width=5cm]{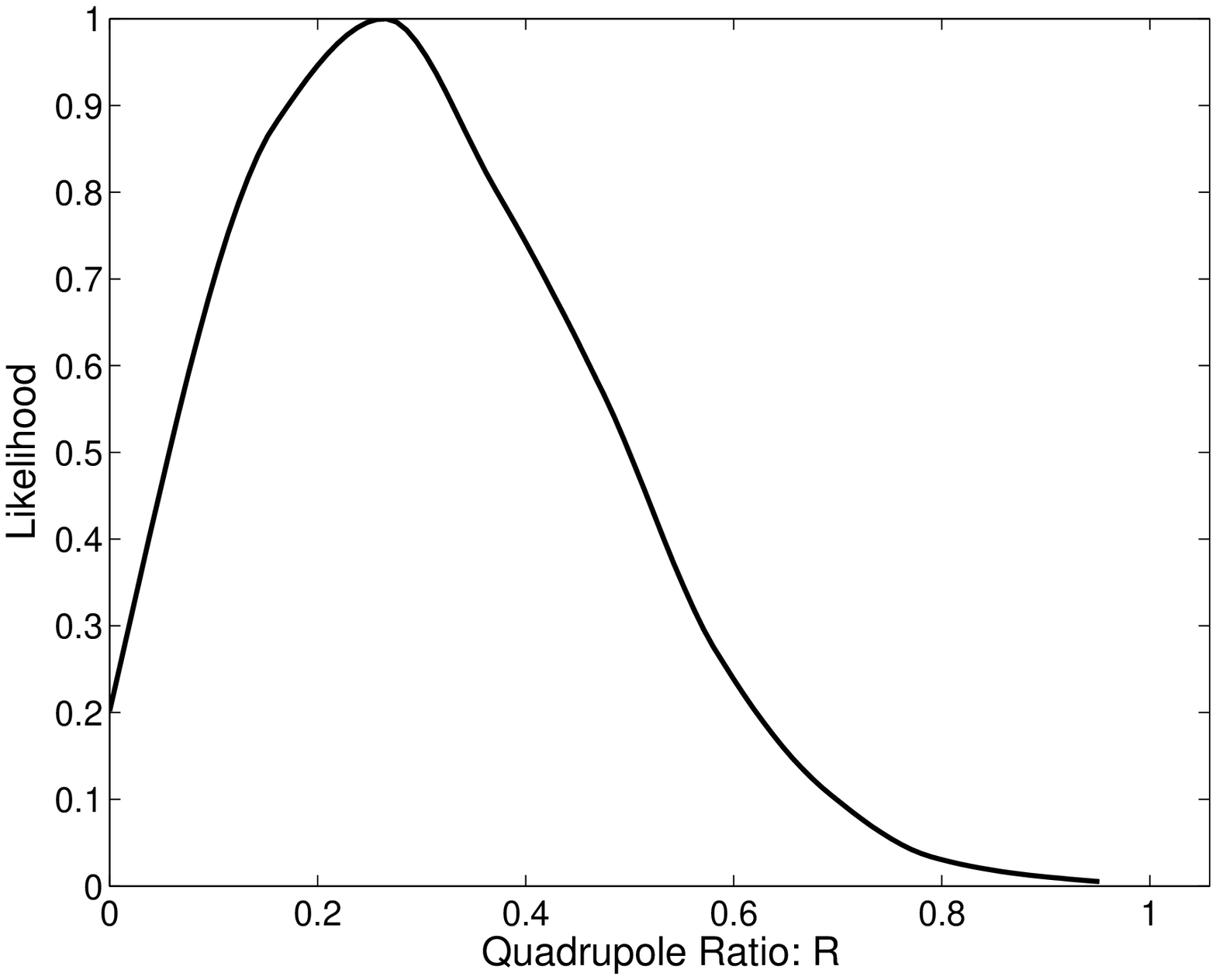}\includegraphics[width=5cm]{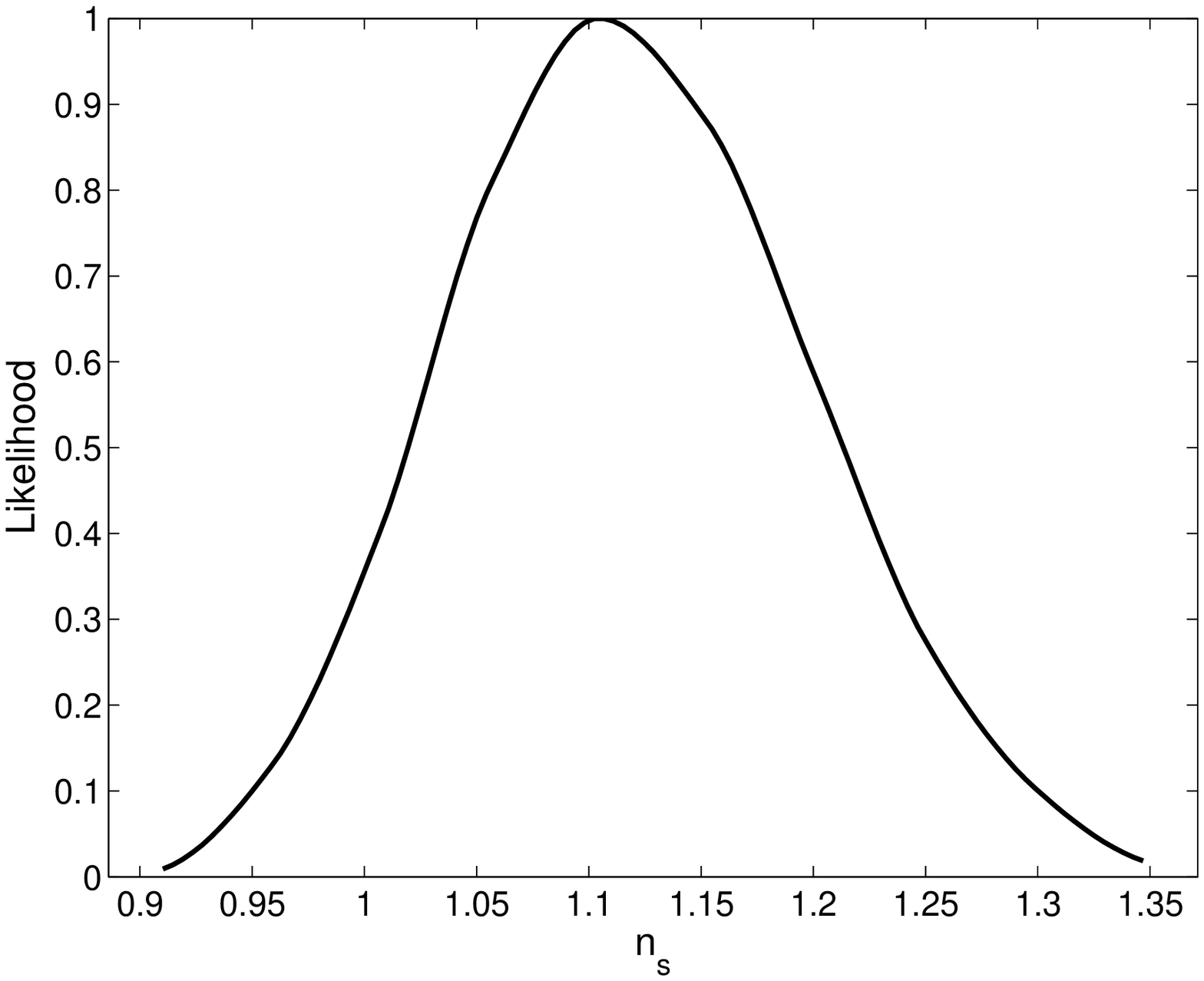}\\ 
\includegraphics[width=5cm]{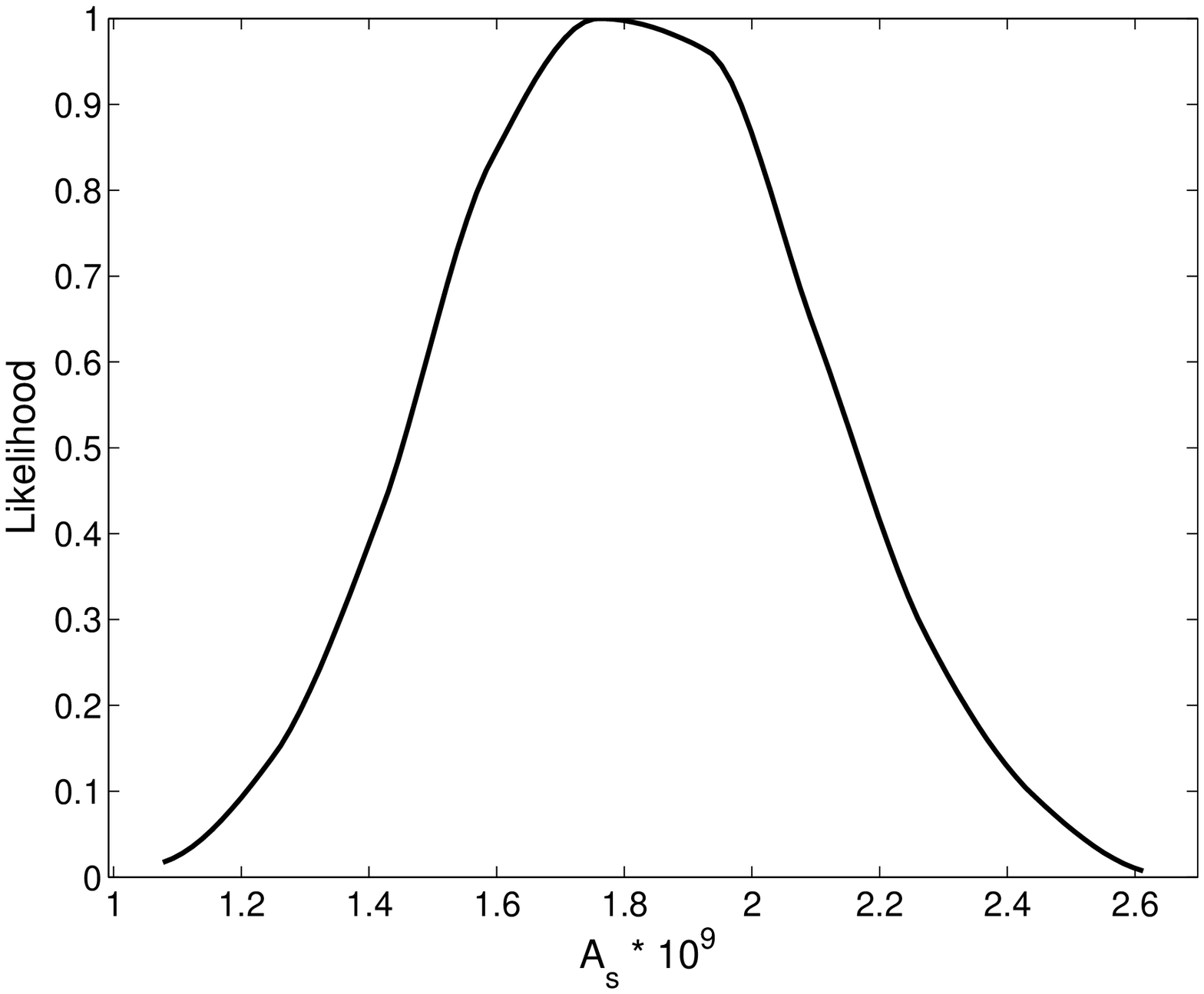}
\end{center}\caption{1-dimensional likelihoods for $R$ (left), $n_s$ (middle) 
and $A_s$ (right). Figure adopted from Zhao et al \cite{zbg2}.}\label{figureb12}
\end{figure}

The marginalized 2-dimensional and 1-dimensional distributions are plotted in 
Fig.~\ref{figurea1} and Fig.~\ref{figureb12}, respectively. The ML values 
for the 1-dimensional marginalized distributions and their $68.3\%$ 
confidence intervals are given by  
\begin{eqnarray}\label{best-fit-1d}
  R=0.266\pm0.171,~~n_s=1.107^{+0.087}_{-0.070} , 
~~A_s=(1.768^{+0.307}_{-0.245})\times10^{-9}.
\end{eqnarray}

%%%%%%%%%%%figure, figure, figure%%%%%%%%%%%%%%%%
%%%%%%%%%%%figure, figure, figure%%%%%

%%%%%%%%%%%figure, figure, figure%%%%%%%%%%%%%%%%
%%%%%%%%%%%figure, figure, figure%%%%%

%%%%%%%%%%%figure, figure, figure%%%%%%%%%%%%%%%%
%%%%%%figure, figure, figure%%%%%

The derived results allow one to conclude that
the maximum likelihood value of $R$ persistently indicates a 
significant amount of relic gravitational waves, even 
if with a considerable uncertainty. The $R=0$ hypothesis (no gravitational 
waves) appears to be away from the $R=0.229$ model at about $1\sigma$ 
interval, or a little more, but not yet at a significantly 
larger distance. The spectral indices $n_s, n_t$ persistently point 
out to the `blue' shape of the primordial spectra, i.e. $n_s >1, n_t >0$, in 
the interval of wavelengths responsible for the analyzed multipoles    
$\ell\le\ell_{max}=100$. This puts in doubt the (conventional) 
scalar fields as a possible driver for the initial kick, because the scalar 
fields cannot support $\beta > -2$ and, consequently, cannot support
$n_s >1, n_t >0$.

%%%%%%%%%%%%%%%%%%%%%%%%%%%%%%%%%%%%%%%%%%%%%%%%%%%%%%%%%%%%%%%%%%%%%%%%%%
%%%%%%%%%%%%%%%%%%%%%%%%%%%%%%%%%%%%%%%%%%%%%%%%%%%%%%%%%%%%%%%%%%%%%%%%%%

\subsection{How relic gravitational waves can be overlooked in the likelihood 
analysis of data\label{section3}}

The results of this analysis differ from the conclusions of WMAP
team \cite{wmap5}. The WMAP team found no evidence for gravitational waves 
and arrived at a `red' spectral index $n_s=0.96$. The 
WMAP findings are symbolized by black dashed and blue dash-dotted contours
in Fig.~\ref{figurea1}. It is important to discuss the likely reasons for 
these disagreements.  

Two main differences in the data analysis are as follows. First, the present 
analysis is restricted only to multipoles $\ell\leq100$ (i.e. to the interval 
in which there is any chance of finding gravitational waves), whereas the WMAP 
team uses the data from all multipoles up to $\ell\sim1000$, keeping 
spectral indices constant in the entire interval of participating 
wavelengths (thus making the uncertainties smaller by increasing 
the number of included data points which have nothing to do with 
gravitational waves). Second, in the current work, the relation $n_t=n_s-1$ 
implied by the theory of quantum-mechanical generation of cosmological 
perturbations is used, whereas
WMAP team uses the inflationary `consistency relation' $r=-8n_t$,
which automatically sends $r$ to zero when $n_t$ approaches zero. 

It is important to realize that the inflationary `consistency relation'
\[
r= 16 \epsilon = - 8 n_t
\]
is a direct consequence of the single contribution of inflationary theory 
to the subject of cosmological perturbations, which is the incorrect 
formula for the power spectrum of density perturbations, 
containing the ``zero in the denominator" factor:
\[ 
P_{s} \approx \frac{1}{\epsilon}\left(\frac{H}{H_{Pl}}\right)^2.
\] 
The ``zero in the denominator" factor $\epsilon$ is 
$\epsilon \equiv - {\dot H}/{H^2}$. This factor tends to zero in the limit
of standard de Sitter inflation ${\dot H}=0$ (any horizontal line in
Fig.~\ref{birth}) independently of the curvature of space-time and strength 
of the generating gravitational field characterized by $H$. To make the wrong 
theory look ``consistent", inflationary model builders push $H/H_{Pl}$ down, 
whenever $\epsilon$ goes to zero (for example, down to the level marked by the 
minutely inclined line ``legitimate transition in inflationary theory" 
in Fig.~\ref{birth}), thus keeping $P_s$ at the observationally required level 
and making the amount of relic gravitational waves arbitrarily small. In fact, 
the most advanced inflationary theories based on strings, branes, 
loops, tori, monodromies, etc. predict the ridiculously small amounts of 
gravitational waves, something at the level of $r \approx 10^{-24}$, or so. 
[There is no doubt, there will be new
inflationary theories which will predict something drastically 
different. Inflationists are good at devising theories that only mother can
love, but not at answering simple physical questions such as quantization
of a cosmological oscillator with variable frequency. For a more detailed
criticism of inflationary theory, see Grishchuk\cite{a12a}.] Obviously,
the analysis in papers \cite{zbg,zbg2} does not use the inflationary theory. 

Baskaran et al concluded that in the conditions of relatively noisy
WMAP data it was the assumed constancy of spectral indices in a broad
spectrum \cite{wmap5} that was mostly responsible for the strong  
dissimilarity of data analysis results with regard to gravitational waves.  
Having repeated the same analysis of data in the adjacent interval of multipoles 
$101 \leq \ell \leq 220$, the authors of paper \cite{zbg2} found a 
somewhat different (smaller) spectral index $n_s$ in this interval. They 
came to the conclusion that the hypothesis of strictly constant spectral indices
(perfectly straight lines in Fig.~\ref{birth}) should be abandoned. It is 
necessary to mention that the constancy of $n_s$ over the vast region of 
wavenumbers, or possibly a simple running of $n_s$, is a usual assumption in a 
number of works \cite{wmap5,othergroups,wishart3}. 

It is now clear why it is dangerous, in the search for relic gravitational 
waves, to include data from higher multipoles, and especially assuming a strictly 
constant spectral index $n_s$. Although the higher multipole data 
have nothing to do with gravitational waves, they bring $n_s$ down. If one 
postulates that $n_s$ is one and the same constant at all $\ell$'s, this 
additional `external' information about $n_s$ affects uncertainty about $R$ 
and brings $R$ down. This is clearly seen, for example, in the left panel of 
Fig.~\ref{figurea1}. The localization of $n_s$ near, say, the line $n_s =0.96$ 
would cross the solid red contours along that line and would enforce lower, or 
zero, values of $R$. However, as was shown \cite{zbg2}, $n_s$ is 
sufficiently different even at the span of the two neighbouring intervals 
of $\ell$'s, namely $2\leq \ell\leq 100$ and $101\leq \ell\leq 220$. These 
considerations, as for how relic gravitational waves can be overlooked in a data
analysis, have general significance and will be applicable to any CMB data.

%%%%%%%%%%%%%%%%%%%%%%%%%%%%%%%%%%%%%%%%%%%%%%%%%%%%%%%%%%%%%%%%%%%%%%%%%%
%%%%%%%%%%%%%%%%%%%%%%%%%%%%%%%%%%%%%%%%%%%%%%%%%%%%%%%%%%%%%%%%%%%%%%%%%%

%%%%%%%%%%%%%%%%%%%%%%%%%%%%%%%%%%%%%%%%%%%%%%%%%%%%%%%%%%%%%%%%%%%%%%%%%%%%
%%%%%%
%%%

\subsection{Forecasts for the Planck mission \label{section4}}

The final part of the presentation by Baskaran et al dealt with 
forecasts for the recently launched Planck satellite \cite{Planck}. 
The ability of a CMB experiment to detect gravitational waves
is characterized by the signal-to-noise ratio defined by
\cite{zbg,zbg2}
\begin{eqnarray}
\label{snr}
 S/N\equiv R/\Delta R,
\end{eqnarray}
where the numerator is the ``true" value of the parameter $R$ (its ML value
or the input value in a numerical simulation) while $\Delta R$ in the
denominator is the expected uncertainty in determination of $R$ from the 
data. 

In formulating the observational expectations, all of the available 
information channels (i.e. $TT$, $TE$, $EE$ and $BB$ correlation functions)  
were taken into account. The calculations were performed using the Fisher matrix 
formalism. The total uncertainty entering the Fisher matrix calculations is 
comprised of instrumental and foreground noises \cite{Planck,cmbpol-fore,zbg2}
as well as the statistical uncertainty of the inherently random signal under
discussion. The possibility of partial removal of contamination arising from foreground sources 
was quantified by the residual noise factor $\sigma^{\rm fg}$. The three considered 
cases included $\sigma^{\rm fg}=1$ (no foreground removal), $\sigma^{\rm fg}=0.1$ 
($10\%$ residual foreground noise) and $\sigma^{\rm fg}=0.01$ ($1\%$ residual 
foreground noise). In order to gauge the worst case scenario, the `pessimistic' 
case was added. It assumes $\sigma^{\rm fg}=1$ and the instrumental noise at 
each frequency $\nu_i$ four times larger than the nominal value reported by the
Planck team. This increased noise is meant to mimic the situation where it is not 
possible to get rid of various systematic effects \cite{systematics}, the $E$-$B$ 
mixing \cite{ebmixture}, cosmic lensing \cite{lensing}, etc. which all affect the 
$BB$ channel.

\begin{figure}
\centerline{\includegraphics[height=8cm]{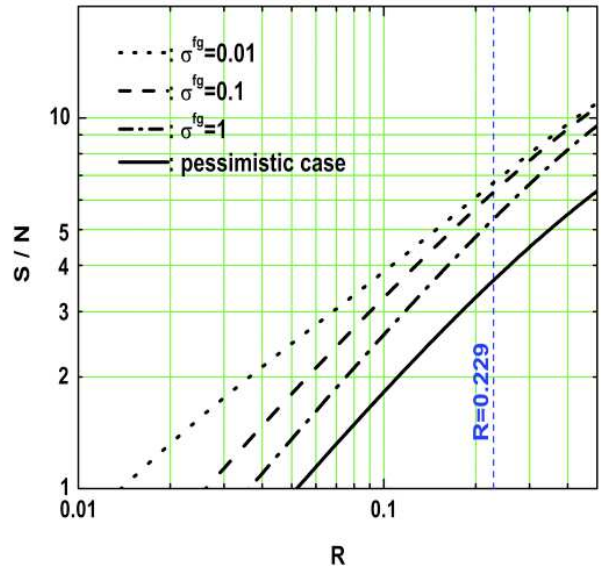}}
\caption{The $S/N$ as a function of $R$. Figure adopted from Zhao et al \cite{zbg2}.}
\label{figurev2}
\end{figure}

The total $S/N$ for one parameter family of models defined by the 
large WMAP5 likelihoods (\ref{1Dmodel}) is plotted in Fig.~\ref{figurev2}. This 
graph is based on the conservative assumption that all unknown parameters 
$R$, $n_s$ and $A_s$ are evaluated from one and the same dataset. 
Four options are depicted: $\sigma^{\rm fg}=0.01,0.1,1$ and the pessimistic 
case. The results for the ML model (\ref{best-fit}) are given by the 
intersection points along the vertical line $R=0.229$. Specifically, 
$S/N = 6.69,~6.20,~5.15$ for $\sigma^{\rm fg}=0.01,~0.1,~1$, respectively. The 
good news is that even in the pessimistic case one gets $S/N>2$ for $R>0.11$, and 
the Planck satellite will be capable of seeing the ML signal $R=0.229$ at 
the level $S/N=3.65$.

%%%%%%%%%%%figure, figure, figure%%%%%%%%%%%%%%%%
%%%%%%%%%%%figure, figure, figure%%%%%

\begin{figure}
\centerline{\includegraphics[width=18cm,height=15cm]{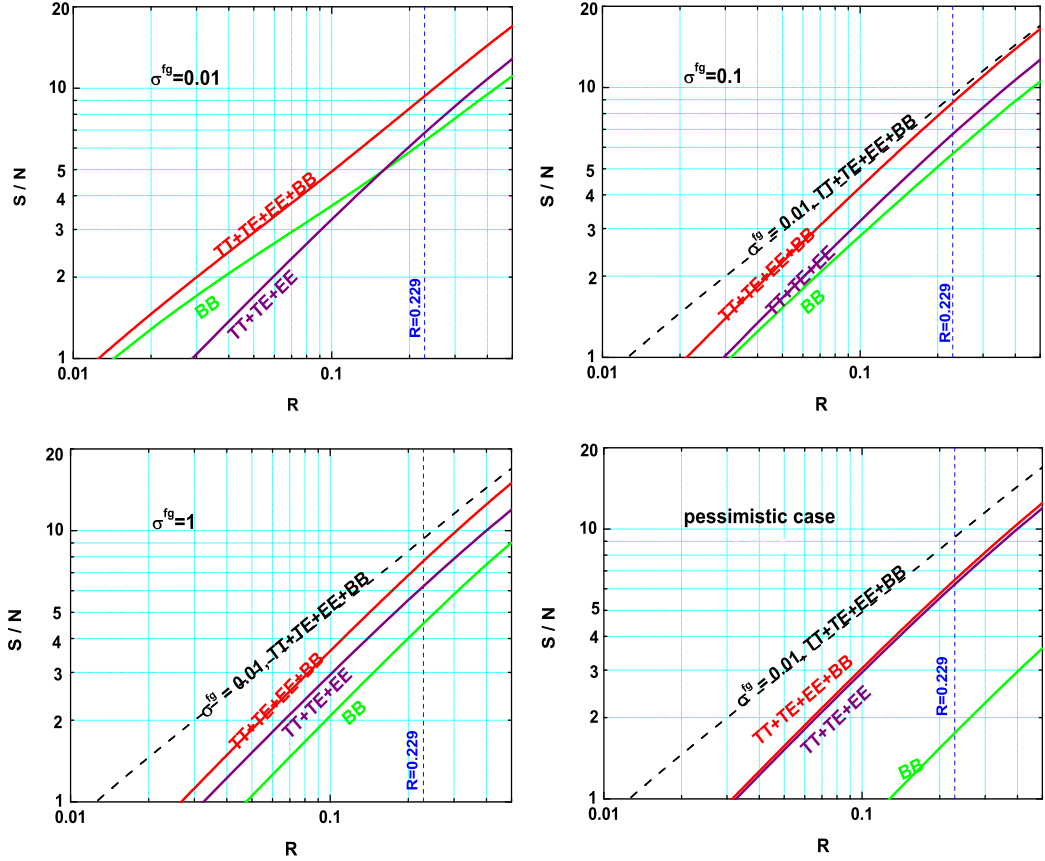}}
\caption{The $S/N$ for different combinations of the information channels, 
$TT+TE+EE+BB$, $TT+TE+EE$ and $BB$. 
Figure adopted from Zhao et al \cite{zbg2}.}\label{figurev11}
\end{figure}

It is important to treat separately the contributions to the total signal-to-noise 
ratio supplied by different information channels and different individual 
multipoles. The $(S/N)^2$ calculated for three combinations 
of channels, $TT+TE+EE+BB$, $TT+TE+EE$ and $BB$ alone, is shown in 
Fig.~\ref{figurev11}. Four panels describe four noise models:
$\sigma^{\rm fg} = 0.01,~0.1,~1$ and the pessimistic case. Surely, the 
combination $TT+TE+EE+BB$ is more sensitive than any of the other two, 
i.e.~$TT+TE+EE$ and $BB$ alone. For example, in the case $\sigma^{\rm fg}=0.1$ 
the use of all correlation functions ensures $S/N$ which is 
$\sim50\%$ greater than $BB$ alone and $\sim30\%$ greater than $TT+TE+EE$. 
The situation is even more dramatic in the pessimistic case. The ML model 
(\ref{best-fit}) can be barely seen through the $B$-modes alone, 
because the $BB$ channel gives only $S/N=1.75$, whereas the use of all 
of the correlation functions can provide a confident detection with $S/N=6.48$. 
Comparing $TT+TE+EE$ with $BB$ one can see that the first method is better 
than the second, except in the case when $\sigma^{\rm fg}=0.01$ and $R$ is 
small ($R<0.16$). In the pessimistic case, the role of the $BB$ channel is so
small that the $TT+TE+EE$ method provides essentially the same sensitivity as 
all channels $TT+TE+EE+BB$ together.

%%%%%%%%%%%figure, figure, figure%%%%%%%%%%%%%%%%
%%%%%%%%%%%figure, figure, figure%%%%%

%%%%%%%%%%%figure, figure, figure%%%%%%%%%%%%%%%%
\begin{figure}
\centerline{\includegraphics[width=18cm,height=15cm]{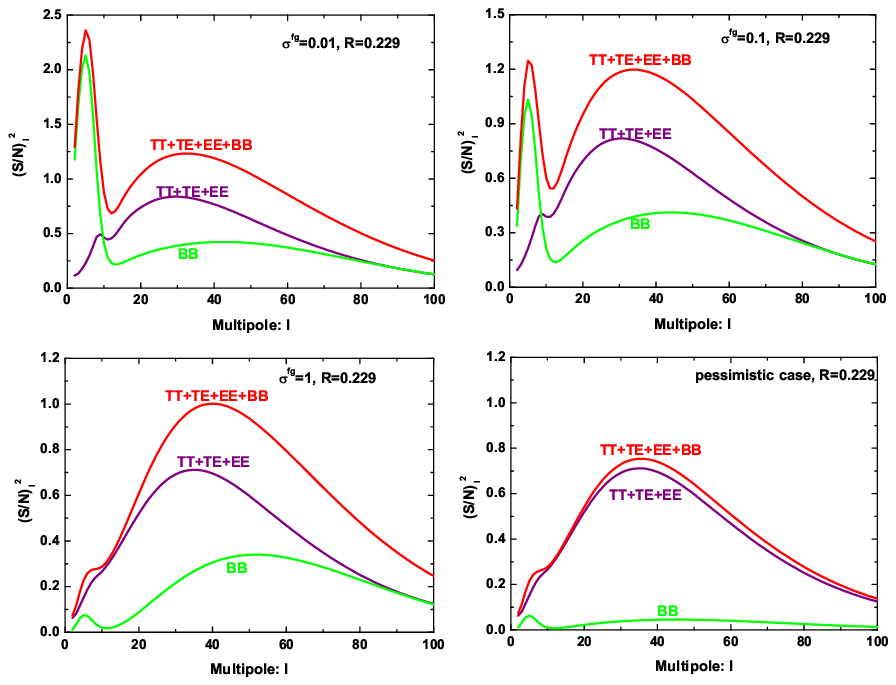}}
\caption{The individual $(S/N)_{\ell}^{2}$ as functions of $\ell$ for 
different combinations of information channels and different
levels of foreground contamination. Calculations are done for the ML
model (\ref{best-fit}) with $R=0.229$.
Figure adopted from Zhao et al \cite{zbg2}.}\label{figurep41}
\end{figure}
%%%%%%%%%%%figure, figure, figure%%%%%

Finally, Fig.~\ref{figurep41} shows the contributions to the 
total signal-to-noise ratio from individual multipoles. 
It can be seen that a very deep foreground cleaning, $\sigma^{\rm fg}=0.01$, 
makes the very low (reionization) multipoles 
$\ell\lesssim 10$ the major contributors to the total $(S/N)^2$, and mostly in
the $BB$ channel. However, for large $\sigma^{\rm fg}=0.1,~1$, and especially in the 
pessimistic case (see the lower right panels in Fig. \ref{figurep41}), the ability of the 
$BB$ channel becomes very degraded at all $\ell$'s. At the same time, as Fig. \ref{figurep41} 
illustrates, the $\ell$-decomposition of $(S/N)^2$ for $TT+TE+EE$ combination depends only 
weakly on the level of $\sigma^{\rm fg}$. Furthermore, in this method, 
the signal to noise curves generally peak at $\ell\sim(20-60)$. Therefore, it 
will be particularly important for Planck mission to avoid excessive 
noises in this region of multipoles.

%%%%%%%%%%%%%%%%%%%%%%%%%%%%%%%%%%%%%%%%%%%%%%%%%%%%%%%%%%%%%%%%%%%%%%%%%%%%%%%%%%
%%%
%%%%%%%%%%%%%%%%%%%%%%%   SECTION 4     %%   SECTION 4  %%   SECTION 4
%%%%%%%%%%%%%%%%%%%%%%%%%%%%%%%%
%%%%%%%%%%%%%%%%%%%%%%%%%%%%%%%%%%%%%%%%%%%%%%%%%%%%%%%%%%%%%%%%%%%%%%%%%%%%%%%%%%

\section{Conclusions}

In general, the OC1 parallel session was balanced and covered
both theoretical and experimental aspects of primordial gravitational waves 
and the CMB. Thanks to excellent contributions of all participants, this subject
received a new momentum. Hopefully, new observations and theoretical work will bring 
conclusive results in the near future.

%%%%%%%%%%%%%%%%%%%%%%%%%%%%%%%%%%%%%%%%%%%%%%%%%%%%%%%%%%%%%%%%%%%%%%%%%%%%%%%%%%%%%%%%%%%%%%%%%%%%%%%%%%%%%%%%%%%%%%%%%%%%%%%%%%%%%%%%%%%%%%%%%%%%%%%%%%%%%%%%%%%

\bibliographystyle{ws-procs975x65}
\bibliography{ws-pro-sample}

\end{document}